\documentclass{article}

\begin{document}

\title{Comments on two papers by Galliano Valent, concerning integrable
Hamiltonian systems admitting quartic and cubic integrals.}
\author{Yehia H.M. \\
Department of Mathematics, Faculty of Science,\\
Mansoura University, Mansoura 35516, Egypt.\\
E-mail: hyehia@mans.edu.eg}
\maketitle

\begin{abstract}
In this note we comment on two recently published papers by G. Valent:

The first is the preprint "On a Class of Integrable Systems with a quartic
First Integral, Arxiv: 1304-5859. April 22, (2013)". We show that the two
integrable Hamiltonian systems introduced in this reprint as original
results are not new. They are special cases of two systems introduced by the
present author in 2006 in two papers \cite{yrcd06} and \cite{ymaster}.

The second paper is "On a Class of Integrable Systems with a Cubic First
Integral, Commun. Math. Phys. 299, 631--649 (2010),

In that paper two integrable hamiltonian systems admitting a cubic integral
were introduced. Those systems were referred to as original results by
Tsiganov in \cite{Tsig11}, Vershilov and Tsiganov in \cite{VerTsig}, Bialy
and Mironov in \cite{BM} and by Gibbons et al in \cite{gib}. We show that
those systems are not new. Both can be obtained as special cases of one
system introduced by us in \cite{ycubic} (2002) and one of them is a special
case of a much earlier version \cite{yint86} published 24 years earlier.
\end{abstract}

\section{\protect\bigskip Introduction}

In the last few decades, a great interest arose in integrable systems on
various types of two-dimensional manifolds. Only few works were devoted to
devise new methods for constructing such systems. The most successful method
seems to be the one introduced by the present author in 1986 \cite{yint86}.
Until now, this method has led to the construction of a large collection of
integrable 2D systems including the richest and biggest ever known ones: 41
time-irreversible systems with a complementary quadratic integral \cite%
{yint92}, \cite{yatlas}; two systems with a cubic integral \cite{yint86}, 
\cite{ycubic} and three systems with an integral of the fourth degree \cite%
{ymaster}, \cite{yrcd06} and \cite{yjpa12}. One of those, called by us
"master", is the biggest ever-known 2D integrable system, in the sense that
it involves 21 free parameters. \bigskip

One of the strong points in our method is the use of Lagrangian formulation
and some point transformations as well as a change of the time
parametrization preserving this formalism. This led to the formation of a
small number of differential equations for the system determination and in
some cases facilitated the process of solution.

Integrable systems constitute a rare exception in hamiltonian dynamics in
general. One must pay due attention to make sure that newly found systems do
not repeat already known ones.\ In the last few years, several trials were
made to construct new integrable 2D systems using Hamiltonian formalism and
(ansatz)es suited for search for certain special forms of quadratic and
cubic integrals. Unfortunately, some authors have rediscovered certain
integrable systems in Hamiltonian form and made no effort to find out that
those systems were obtained significantly earlier, but in some other forms.
Some authors give either incomplete or erroneous references; and other
authors do not provide due references, leaving the reader with the
impression that provided results are new. Some of the most notable examples
are in the works \cite{valcmp}-\cite{Val-Arx}.

\section{Two systems with a quartic integral in \protect\bigskip Valent's
paper \protect\cite{Val-Arx}}

In a recent work \cite{Val-Arx} Valent introduced two integrable Hamiltonian
systems which, in his words, "generalize, to some extent, the results on
integrable geodesic flows on two dimensional manifolds with a quartic first
integral in the framework laid down by Selivanova and Hadeler".

\bigskip As a motivation of the method used by him in \cite{valcmp} and \cite%
{Val-Arx} Valent says "a more direct analysis of the differential system
leading to integrability revealed to be also successful:". In fact, this is
the same method used in our earlier works including \cite{yint86} and, in
particular, in \cite{yrcd06}, where it has led to more general systems than
those given in \cite{Val-Arx}. Apart from this, Valent's conclusions about
restoring integrable cases of rigid body motions are not only repetitions of
earlier results, but also very special cases of them.

\subsection{\protect\bigskip The first system of \protect\cite{Val-Arx}}

That is given as a Hamiltonian system in theorem 2 of \cite{Val-Arx}. Its
Lagrangian equivalent is characterized by the Lagrangian

\begin{eqnarray}
L &=&\frac{1}{2}\beta ^{2}(\frac{\dot{x}^{2}}{F}+\frac{\dot{\varphi}^{2}}{%
\sqrt{F}})  \nonumber \\
&&-\frac{1}{2\beta ^{2}}(k\sqrt{F}G^{\prime }(x)\cos \varphi +lG(x)+mx+n)
\label{L1v}
\end{eqnarray}%
where 
\begin{eqnarray*}
\beta ^{2} &=&b_{0}+\alpha x, \\
F &=&x^{4}+c_{2}x^{2}+c_{1}x+c_{0}, \\
G(x) &=&\sqrt{F}-x^{2}-\frac{c_{2}}{2}
\end{eqnarray*}%
It can be easily verified that this system is a special case of the system
introduced in 2006 in our work \cite{yrcd06} with the Lagrangian

\begin{eqnarray}
L &=&\frac{1}{2}[(\frac{\mu \alpha _{2}p^{2}+\alpha _{1}p+\alpha _{0}}{\sqrt{%
\mu ^{2}p^{4}+c_{2}p^{2}+c_{1}p+c_{0}}}-\alpha _{2})\dot{\xi}^{2}  \nonumber
\\
&&\ \ \ \ \ \ +\ \frac{(\mu \alpha _{2}p^{2}+\alpha _{1}p+\alpha _{0}-\alpha
_{2}\sqrt{\mu ^{2}p^{4}+c_{2}p^{2}+c_{1}p+c_{0}})}{\mu
^{2}p^{4}+c_{2}p^{2}+c_{1}p+c_{0}}\dot{p}^{2}]  \nonumber \\
&&+\frac{\mu h_{3}p^{2}-h_{1}p-h_{2}}{\sqrt{\mu
^{2}p^{4}+c_{2}p^{2}+c_{1}p+c_{0}}}  \nonumber \\
&&-\frac{-\mu h_{3}p^{2}+h_{1}p+h_{2}-h_{3}\sqrt{\mu
^{2}p^{4}+c_{2}p^{2}+c_{1}p+c_{0}}}{\mu \alpha _{2}p^{2}+\alpha _{1}p+\alpha
_{0}-\alpha _{2}\sqrt{\mu ^{2}p^{4}+c_{2}p^{2}+c_{1}p+c_{0}}}  \nonumber \\
&&+\frac{[4\mu ^{2}p^{3}+2c_{2}p+c_{1}-4\mu p\sqrt{\mu
^{2}p^{4}+c_{2}p^{2}+c_{1}p+c_{0}}]}{16\mu (\mu \alpha _{2}p^{2}+\alpha
_{1}p+\alpha _{0}-\alpha _{2}\sqrt{\mu ^{2}p^{4}+c_{2}p^{2}+c_{1}p+c_{0}})}%
A\cos (\sqrt{\mu }\xi )+h  \label{L1}
\end{eqnarray}%
From this Lagrangian we can get (\ref{L1v}) simply by enforcing the
conditions%
\begin{eqnarray}
\mu &=&1,A=-16k,  \nonumber \\
\alpha _{0} &=&\alpha ,\alpha _{1}=b_{0},\alpha _{2}=0,  \nonumber \\
h_{1} &=&2m,h_{2}=2n-lc_{2},h_{3}=2l
\end{eqnarray}%
and renaming the variables $p\rightarrow x,\xi \rightarrow \varphi .$

Valent's Lagrangian (\ref{L1v}) contains only 6 free parameters, i.e. two
parameters less than our Lagrangian (\ref{L1}) involving 8 parameters.

Our paper \cite{yrcd06} containing (\ref{L1}) and published 7 years earlier
than \cite{Val-Arx} is not mentioned at all in \cite{Val-Arx}.

\subsection{\protect\bigskip The second system of \protect\cite{Val-Arx}}

\bigskip That is given as a Hamiltonian system in theorem 3 of \cite{Val-Arx}%
. Its Lagrangian equivalent is characterized by the Lagrangian%
\begin{eqnarray}
L &=&\frac{1}{2r}(-1+\frac{x^{2}+d}{\sqrt{R}})(\frac{\dot{x}^{2}}{\sqrt{R}}+%
\dot{\varphi}^{2})  \nonumber \\
&&-\frac{1}{2\beta ^{2}}(k\sqrt{F}\alpha \cos \varphi +lG(x)+mx+n)
\label{L2v}
\end{eqnarray}%
where 
\begin{eqnarray*}
F &=&x^{4}-4rx^{3}\alpha +2(d+2r^{2}\alpha ^{2}-2rc_{2})x^{2}-4\allowbreak
c_{1}rx+d^{2}-4c_{0}r, \\
G &=&\alpha x+c_{2}-\beta ^{2}, \\
\beta ^{2} &=&\frac{(x^{2}+d)-\sqrt{F}}{2r}
\end{eqnarray*}

This system is obtained from our "master" system of \cite{ymaster}, whose
21-parameter Lagrangian is

\begin{eqnarray}
L &=&\frac{1}{2}\Lambda \lbrack \frac{\dot{p}^{2}}{\sqrt{%
a_{4}p^{4}+a_{3}p^{3}+a_{2}p^{2}+a_{1}p+a_{0}}}+\frac{\dot{q}^{2}}{\sqrt{%
a_{4}q^{4}+b_{3}q^{3}+b_{2}q^{2}+b_{1}q+b_{0}}}]  \nonumber \\
&&-\frac{1}{\Lambda }\{[\frac{h_{0}b_{3}p^{3}+h_{3}p^{2}+h_{2}p+h_{1}}{\sqrt{%
a_{4}p^{4}+a_{3}p^{3}+a_{2}p^{2}+a_{1}p+a_{0}}}+\frac{%
h_{0}a_{3}q^{3}+h_{3}q^{2}+h_{5}q+h_{4}}{\sqrt{%
a_{4}q^{4}+b_{3}q^{3}+b_{2}q^{2}+b_{1}q+b_{0}}}]  \nonumber \\
&&+h_{0}[\frac{q(4a_{4}p^{3}+3a_{3}p^{2}+2a_{2}p+a_{1})}{\sqrt{%
a_{4}p^{4}+a_{3}p^{3}+a_{2}p^{2}+a_{1}p+a_{0}}}  \nonumber \\
&&+\frac{p(4a_{4}q^{3}+3b_{3}q^{2}+2b_{2}q+b_{1})}{\sqrt{%
a_{4}q^{4}+b_{3}q^{3}+b_{2}q^{2}+b_{1}q+b_{0}}}]\}  \label{LT}
\end{eqnarray}%
where%
\begin{eqnarray*}
\Lambda &=&[\frac{\alpha _{0}b_{3}p^{3}+\alpha _{3}p^{2}+\alpha _{2}p+\alpha
_{1}}{\sqrt{a_{4}p^{4}+a_{3}p^{3}+a_{2}p^{2}+a_{1}p+a_{0}}}+\frac{\alpha
_{0}a_{3}q^{3}+\alpha _{3}q^{2}+\alpha _{5}q+\alpha _{4}}{\sqrt{%
a_{4}q^{4}+b_{3}q^{3}+b_{2}q^{2}+b_{1}q+b_{0}}}] \\
&&+\alpha _{0}[\frac{q(4a_{4}p^{3}+3a_{3}p^{2}+2a_{2}p+a_{1})}{\sqrt{%
a_{4}p^{4}+a_{3}p^{3}+a_{2}p^{2}+a_{1}p+a_{0}}}+\frac{%
p(4a_{4}q^{3}+3b_{3}q^{2}+2b_{2}q+b_{1})}{\sqrt{%
a_{4}q^{4}+b_{3}q^{3}+b_{2}q^{2}+b_{1}q+b_{0}}}]
\end{eqnarray*}%
by imposing on the last system the conditions%
\begin{eqnarray}
a_{4} &=&1,b_{2}=-2,b_{0}=1,b_{3}=b_{1}=0,  \nonumber \\
a_{0} &=&-4rc_{0}+d^{2},a1=-4c_{1}r,a_{3}=-4r\alpha ,  \nonumber \\
a_{2} &=&-4c_{2}r+2d+4\alpha ^{2}r^{2},  \nonumber \\
\alpha _{0} &=&\alpha _{2}=\alpha _{5}=0,\alpha _{1}=\frac{d}{r},\alpha _{3}=%
\frac{1}{r}=-\alpha _{4},  \nonumber \\
h_{0} &=&\frac{k}{2r},h_{1}=2n+2lc_{2}-\frac{ld}{r},  \nonumber \\
h_{2} &=&2m+2l\alpha ,-h_{3}=h_{4}=\frac{1}{r},h_{5}=2\alpha k
\end{eqnarray}%
and changing the variables $p\rightarrow x,q\rightarrow \cos \varphi .$ That
is a restriction of the 21-parameter system to a 10-parameters one. As this
case is the most involved, we provide in the appendix a MAPLE code which can
be used to verify the conclusion of the present subsection concerning the
Lagrangians (\ref{L2v}) and(\ref{LT}).

It was not mentioned at all in (\cite{Val-Arx}) that the system in theorem 3
of that work is a special case of (\ref{LT}) published 7 years earlier in 
\cite{ymaster}.

\section{Two systems with a cubic integral in \protect\bigskip 
{\protect\LARGE Valent's paper }\protect\cite{valcmp}}

\bigskip \textit{The content of this section was submitted to
"Communications in Mathematical Physics" as a comment. After reviewing the
comment, the Editor decided that it is sufficient to publish the Erratum 
\cite{val-erratum}, acknowledging the situation without refering to our
comment. As the same situation is now repeated, we provide the full comment
below.}

In this section we consider in detail the two systems with a cubic integral
pointed out by Valent in \cite{valcmp} as new findings. Those systems were
further refered to as original results in \cite{VerTsig}, \cite{BM} and \cite%
{gib}. We show that both systems are special cases of one of the systems
announced much earlier in our work \cite{ycubic} and one of them is
equivalent to a special case constructed 24 years earlier in \cite{yint86}.
Works like \cite{valcmp} (as well as \cite{val-yah}, which was briefly
commented in \cite{yatlas} and \cite{tsiglet} commented in detail in \cite%
{ycom1}), have brought some confusion concerning literature in the field of
integrable systems, and we feel some light must be shed on the present,
seemingly chaotic situation created by the commented paper and other works
citing it.

{\LARGE Valent's systems and their Lagrangian analogs}

The Hamiltonian characterizing the first system of \cite{valcmp}, which we
denote by $H_{v1}$, can be explicitly expressed using relations (14) and(15)
of that work as%
\begin{eqnarray}
H_{v1} &=&\frac{1}{2}[(\zeta ^{3}+3c_{0}\zeta -2\rho _{0})p_{\zeta }^{2}+%
\frac{-3\zeta ^{4}-18c_{0}\zeta ^{2}+24\rho _{0}\zeta +9c_{0}^{2}}{4(\zeta
^{3}+3c_{0}\zeta -2\rho _{0})}p_{\varphi }^{2}]  \nonumber \\
&&+\lambda \sqrt{\zeta ^{3}+3c_{0}\zeta -2\rho _{0}}cos\varphi +m\zeta
\label{Hv1}
\end{eqnarray}%
The corresponding Lagrangian, which we denote by $L_{v1}$, in the same
generalized coordinates $\zeta ,\varphi $ is%
\begin{eqnarray}
L_{v1} &=&\frac{1}{2}[\frac{\dot{\zeta}^{2}}{\zeta ^{3}+3c_{0}\zeta -2\rho
_{0}}+\frac{4(\zeta ^{3}+3c_{0}\zeta -2\rho _{0})}{3(-3\zeta
^{4}-18c_{0}\zeta ^{2}+24\rho _{0}\zeta +9c_{0}^{2})}\dot{\varphi}^{2}] 
\nonumber \\
&&-\lambda \sqrt{\zeta ^{3}+3c_{0}\zeta -2\rho _{0}}cos\varphi -m\zeta
\label{Lv1}
\end{eqnarray}

\bigskip The second system in Valent's work \cite{valcmp}, as follows from
formulas (39)-(40) of that work, has the Hamiltonian 
\begin{eqnarray}
H_{v2} &=&\frac{1}{2\varsigma }[(c_{3}\zeta ^{3}+c_{2}\zeta ^{2}+c_{1}\zeta
+c_{0})p_{\zeta }^{2}  \nonumber \\
&&+\frac{-3c_{3}^{2}\zeta ^{4}-4c_{2}c_{3}\zeta ^{3}-6c_{1}c_{3}\zeta
^{2}-12c_{0}c_{3}\zeta +c_{1}^{2}-4c_{0}c_{2}}{4(c_{3}\zeta ^{3}+c_{2}\zeta
^{2}+c_{1}\zeta +c_{0})}p_{\varphi }^{2}]  \nonumber \\
&&+\frac{1}{2q\varsigma }(\sqrt{c_{3}\zeta ^{3}+c_{2}\zeta ^{2}+c_{1}\zeta
+c_{0}}cos\varphi +\beta _{0})  \label{Hv2}
\end{eqnarray}%
The corresponding Lagrangian, which we denote by $L_{v2}$, in the same
generalized coordinates $\zeta ,\varphi $ is%
\begin{eqnarray}
L_{v2} &=&\frac{1}{2}[\frac{\zeta \dot{\zeta}^{2}}{c_{3}\zeta
^{3}+c_{2}\zeta ^{2}+c_{1}\zeta +c_{0}}  \nonumber \\
&&+\frac{4\zeta (c_{3}\zeta ^{3}+c_{2}\zeta ^{2}+c_{1}\zeta +c_{0})\dot{%
\varphi}^{2}}{-3c_{3}^{2}\zeta ^{4}-4c_{2}c_{3}\varsigma
^{3}-6c_{1}c_{3}\zeta ^{2}-12c_{0}c_{3}\zeta +c_{1}^{2}-4c_{0}c_{2}}] 
\nonumber \\
&&-\frac{1}{2q\zeta }(\sqrt{c_{3}\zeta ^{3}+c_{2}\zeta ^{2}+c_{1}\zeta +c_{0}%
}cos\varphi +\beta _{0})  \label{Lv2}
\end{eqnarray}

\bigskip {\LARGE Our system }

The related system in our work \cite{ycubic} is expressed by the formula
(81) of that work%
\begin{eqnarray}
L_{y} &=&\frac{1}{2}[\frac{3(\gamma \nu +\delta )}{\mu (4\nu ^{3}+6\alpha
\nu -\beta )}(\frac{d\nu }{dt})^{2}+\frac{(\gamma \nu +\delta )(4\nu
^{3}+6\alpha \nu -\beta )}{(3\alpha ^{2}+4\beta \nu -12\alpha \nu ^{2}-4\nu
^{4})}(\frac{d\varphi }{dt})^{2}]  \nonumber \\
&&+\frac{C(2\nu ^{2}+\alpha )+D(4\nu ^{3}-6\alpha \nu +2\beta )}{(3\alpha
^{2}+4\beta \nu -12\alpha \nu ^{2}-4\nu ^{4})}\frac{d\varphi }{dt}  \nonumber
\\
&&+\frac{1}{3(\gamma \nu +\delta )}[3(a\nu +b)+\frac{2C^{2}\nu +6CD(2\nu
^{2}-\alpha )+6D^{2}(4\nu ^{3}-\beta )}{3\alpha ^{2}+4\beta \nu -12\alpha
\nu ^{2}-4\nu ^{4}}  \nonumber \\
&&\qquad \qquad \qquad +4\sqrt{4\nu ^{3}+6\alpha \nu -\beta }\Phi (\varphi )]
\label{Ly}
\end{eqnarray}

where 
\begin{equation}
\Phi (\varphi )={\Huge \{}%
\begin{array}{ll}
A\cos \sqrt{\mu }(\varphi -\varphi _{0}) & \mu >0 \\ 
Ae^{\sqrt{-\mu }\varphi }+Be^{-\sqrt{-\mu }\varphi } & \mu <0%
\end{array}
\label{fi}
\end{equation}

Specially notable is the presence in $L_{y}$ of the extra-parameters $C,D\ $%
\ and $\mu $. $C,D$ invoke terms linear in the velocities and make our
system time-irreversible. The parameter $\mu $ widens the range of
applications of the system to special cases, for example, in different
integrable dynamics of the rigid body. Negative values of $\mu $
characterize certain integrable particle dynamics of the Toda type \cite%
{ycubic}.

We state the following notices on Valent's work:

\begin{enumerate}
\item \emph{Both systems of Valent characterized by (\ref{Hv1}, \ref{Lv1})
and (\ref{Hv2}, \ref{Lv2}) are special cases of our system characterized by
the 11-parameter Lagrangian (\ref{Ly}) .}
\end{enumerate}

\bigskip In fact, the 4-parameter Lagrangian (\ref{Lv1}) can be\ obtained
from (\ref{Ly}) by\ enforcing the following restrictions on 7 parameters 
\begin{equation}
C=D=\gamma =b=\varphi _{0}=0,\delta =4/3,\mu =1
\end{equation}%
and renaming the remaining 4 parameters%
\begin{equation}
\alpha =2c_{0},\beta =8\rho _{0},A=-\lambda /16,a=-4m/3
\end{equation}%
\bigskip \bigskip Similarly, (\ref{Lv2}) can be readily obtained from (\ref%
{Ly}) by imposing the conditions $C=D=a=\varphi _{0}=0,$ $\mu =\gamma =1,$
introducing a change of the coordinate $v=\zeta -\delta ,$ and renaming the
parameters 
\begin{eqnarray}
\alpha &=&\frac{2(3c_{1}c_{3}-c_{2}^{2})}{9c_{3}^{2}},\beta =-\frac{%
4(2c_{2}^{3}+27c_{0}c_{3}^{2}-9c_{1}c_{2}c_{3})}{27c_{3}^{3}},  \nonumber \\
\delta &=&-\frac{c_{2}}{c_{3}},A=-\frac{9c_{3}}{64q},b=-\frac{2\beta _{0}}{2q%
}
\end{eqnarray}

It is worth mentioning that the two systems (\ref{Hv1},\ref{Hv2}) were
treated in \cite{valcmp} separately. In fact, Valent states at the end of \S %
2 of \cite{valcmp} that "The special case $q=0$ is rather difficult to
obtain as the limit of the general case $q\neq 0$, so we will first work it
out completely".

\begin{enumerate}
\item[2] A limited version of the Lagrangian (\ref{Ly}) obtained 24 years
earlier than the publication of \cite{valcmp} may be expressed from formulas
(33) of \cite{yint86}, after the usual time change, as%
\begin{eqnarray}
L_{y0} &=&\frac{1}{2}[\frac{\dot{\eta}^{2}}{\mu (\eta ^{3}+12\alpha \eta -12)%
}+\frac{4(\eta ^{3}+12\alpha \eta -12)\dot{\varphi}^{2}}{3(48\alpha
^{2}+48\eta -24\alpha \eta ^{2}-\eta ^{4})}]  \nonumber \\
&&+\frac{\beta }{3}\sqrt{\eta ^{3}+12\alpha \eta -12}cos(\sqrt{\mu }\varphi
)+\frac{p_{0}}{2}\eta  \label{Ly86}
\end{eqnarray}
\end{enumerate}

It is not hard to verify that the Lagrangian $L_{v1}$ of Valent's first
system can be obtained from (\ref{Ly86}) by setting $\mu =1,$ rescaling the
generalized coordinate $\eta $ in (\ref{Ly86}) by the relation 
\begin{equation}
\eta =\sqrt[3]{\frac{6}{\rho _{0}}}\varsigma
\end{equation}%
and renaming the parameters according to the formulas%
\begin{equation}
\alpha =\sqrt[3]{\frac{9}{16\rho _{0}^{2}}}c_{0},\beta =-\frac{3\lambda }{8}%
\sqrt[6]{6\ast \rho _{0}^{5}},p_{0}=-\frac{m}{4}\sqrt[3]{6\ast \rho _{0}^{2}}
\end{equation}

\begin{enumerate}
\item[4] The equation (26) of Valent's commented paper attributed to
Selivanova \cite{sel} is the special case ($\mu =1$) of equation (28) of our
1986 paper \cite{yint86}. The solution of this equation in the form (27) of 
\cite{valcmp}, claimed by Valent to be new, is exactly the same as the
solution given, 24 years earlier, by the first equations of (33) and (34) of 
\cite{yint86} for $\mu =1.$ The same equation with more details on the
process of solution is given in \cite{ycubic}, where also Selivanova's work
was commented in \S 3.

\item[5] In the last page of the commented paper, while he counts known
related integrable systems, the author of \cite{valcmp} mentions
Goriachev-Chaplygin's, Dullin-Matveev's and Goriachev's cases, but without
reference what soever for the first and third cases. In fact, both those
cases were obtained for the first time as special cases of a more general
integrable system admitting a cubic integral in our 1986 work \cite{yint86}.
In our later paper \cite{ycubic} both systems were significantly generalized
by adding several parameters invoking additional potential and gyroscopic
forces.

\item[6] In spite of the facts that:

\begin{enumerate}
\item our works \cite{ycubic} and \cite{yint86} were published 8 and 24
years earlier than the commented article \cite{valcmp},

\item our works \cite{ycubic} and \cite{yint86} contain essential results of 
\cite{valcmp} as special cases,

\item the authors of the work \cite{dulmat}\bigskip , which is\ cited in 
\cite{valcmp} as ref. 4, devote a special section (\S 4. Remark) at the end
of\ their article to acknowledge that the integrable system studied by them
is contained in our results in \cite{ycubic},
\end{enumerate}
\end{enumerate}

neither of our works is mentioned in \cite{valcmp}.

\section{\protect\bigskip Appendix}

The MAPLE code below is used to directly and explicitly show how to get the
second case of Valent's work \cite{Val-Arx} (see section 2.2 above) as
special case of our MASTER system:

Lv := -1/2*ft\symbol{94}2*((-x\symbol{94}2-d)*(x\symbol{94}4-4*r*alpha*x%
\symbol{94}3+(-4*c2*r+2*d+4*alpha\symbol{94}2*r\symbol{94}2)*x\symbol{94}%
2-4*r*c1*x+d\symbol{94}2-4*r*c0)\symbol{94}(1/2)+x\symbol{94}4-4*r*alpha*x%
\symbol{94}3

+(-4*c2*r+2*d+4*alpha\symbol{94}2*r\symbol{94}2)*x\symbol{94}2-4*r*c1*x+d%
\symbol{94}2-4*r*c0)/(x\symbol{94}4-4*r*alpha*x\symbol{94}%
3+(-4*c2*r+2*d+4*alpha\symbol{94}2*r\symbol{94}2)*x\symbol{94}2-4*r*c1*x+d%
\symbol{94}2-4*r*c0)/r

+1/2*xt\symbol{94}2*(-(x\symbol{94}4-4*r*alpha*x\symbol{94}%
3+(-4*c2*r+2*d+4*alpha\symbol{94}2*r\symbol{94}2)*x\symbol{94}2-4*r*c1*x+d%
\symbol{94}2-4*r*c0)\symbol{94}(1/2)+x\symbol{94}2+d)/(4*x\symbol{94}2*r%
\symbol{94}2*alpha\symbol{94}2

+(-4*alpha*x\symbol{94}3-4*x\symbol{94}2*c2-4*c1*x-4*c0)*r+(x\symbol{94}2+d)%
\symbol{94}2)/r+(((-2*x\symbol{94}3*r*alpha\symbol{94}%
2+(c1-d*alpha-2*c2*alpha*r+2*alpha\symbol{94}3*r\symbol{94}2)*x\symbol{94}2

+((-2*c1*alpha+2*d*alpha\symbol{94}%
2)*r+2*c0-2*d*c2)*x-2*r*c0*alpha+d*(-c1+d*alpha))*k*cos(phi)+m*x\symbol{94}%
3+(n+l*r*alpha\symbol{94}2)*x\symbol{94}2

+(-l*c1+l*alpha*d+m*d)*x+(n+l*c2)*d-l*c0)*(x\symbol{94}4-4*r*alpha*x\symbol{%
94}3+(-4*c2*r+2*d+4*alpha\symbol{94}2*r\symbol{94}2)*x\symbol{94}2-4*r*c1*x+d%
\symbol{94}2-4*r*c0)\symbol{94}(1/2)

+(2*alpha\symbol{94}2*x\symbol{94}5*r+(6*c2*alpha*r-d*alpha-6*alpha\symbol{94%
}3*r\symbol{94}2+c1)*x\symbol{94}4+(4*alpha\symbol{94}4*r\symbol{94}3-8*alpha%
\symbol{94}2*c2*r\symbol{94}2+(-2*d*alpha\symbol{94}2+4*c2\symbol{94}%
2+4*c1*alpha)*r

+2*c0-2*d*c2)*x\symbol{94}3+4*(alpha\symbol{94}%
2*(d*alpha-3/2*c1)*r+(1/2*c0-d*c2)*alpha+3/2*c2*c1)*r*x\symbol{94}2+(-4*c0*r%
\symbol{94}2*alpha\symbol{94}2+(-4*alpha*d*c1+4*c0*c2

+2*d\symbol{94}2*alpha\symbol{94}2+2*c1\symbol{94}2)*r-2*d\symbol{94}%
2*c2+2*d*c0)*x-4*c0*(-1/2*c1+d*alpha)*r+d\symbol{94}%
2*(-c1+d*alpha))*k*cos(phi)+m*x\symbol{94}5

+((-l*alpha\symbol{94}2-2*m*alpha)*r+n)*x\symbol{94}4+((2*l*alpha\symbol{94}%
3+2*m*alpha\symbol{94}2)*r\symbol{94}%
2+((-2*n-4*l*c2)*alpha-2*m*c2)*r+2*m*d-l*c1+l*alpha*d)*x\symbol{94}3

+(2*alpha\symbol{94}2*(n+l*c2)*r\symbol{94}2+(-2*n*c2+l*alpha\symbol{94}%
2*d-2*l*c2\symbol{94}2-2*l*alpha*c1-2*c1*m)*r+(2*n+l*c2)*d-l*c0)*x\symbol{94}%
2

+((-2*m*c0-2*n*c1-2*l*alpha*c0-2*l*c2*c1)*r+d*(-l*c1+l*alpha*d+m*d))*x-2*c0*(n+l*c2)*r+((n+l*c2)*d-l*c0)*d)/(x%
\symbol{94}2+d+(x\symbol{94}4-4*r*alpha*x\symbol{94}3

+(-4*c2*r+2*d+4*alpha\symbol{94}2*r\symbol{94}2)*x\symbol{94}2-4*r*c1*x+d%
\symbol{94}2-4*r*c0)\symbol{94}(1/2))/(-alpha*x\symbol{94}3+(-c2+r*alpha%
\symbol{94}2)*x\symbol{94}2-c1*x-c0):

L := 1/2*((p\symbol{94}3*b3*A0+p\symbol{94}2*A3+A1+A2*p)/(sqrt(a4*p\symbol{94%
}4+a3*p\symbol{94}3+a2*p\symbol{94}2+a1*p+a0))+(A0*q\symbol{94}3*a3+A3*q%
\symbol{94}2+A4+A5*q)

/(sqrt(a4*q\symbol{94}4+b3*q\symbol{94}3+b2*q\symbol{94}%
2+b1*q+b0))+A0*(4*a4*p\symbol{94}3+3*a3*p\symbol{94}2+2*a2*p+a1)*q/(sqrt(a4*p%
\symbol{94}4+a3*p\symbol{94}3+a2*p\symbol{94}2+a1*p+a0))

+A0*p*(4*a4*q\symbol{94}3+3*b3*q\symbol{94}2+2*b2*q+b1)/(sqrt(a4*q\symbol{94}%
4+b3*q\symbol{94}3+b2*q\symbol{94}2+b1*q+b0)))*(pt\symbol{94}2/(sqrt(a4*p%
\symbol{94}4+a3*p\symbol{94}3+a2*p\symbol{94}2+a1*p+a0))

+qt\symbol{94}2/(sqrt(a4*q\symbol{94}4+b3*q\symbol{94}3+b2*q\symbol{94}%
2+b1*q+b0)))-((p\symbol{94}3*b3*h0+p\symbol{94}2*h3+h1+h2*p)/(sqrt(a4*p%
\symbol{94}4+a3*p\symbol{94}3+a2*p\symbol{94}2+a1*p+a0))+(h0*q\symbol{94}3*a3

+h3*q\symbol{94}2+h4+h5*q)/(sqrt(a4*q\symbol{94}4+b3*q\symbol{94}3+b2*q%
\symbol{94}2+b1*q+b0))+h0*(4*a4*p\symbol{94}3+3*a3*p\symbol{94}%
2+2*a2*p+a1)*q/(sqrt(a4*p\symbol{94}4+a3*p\symbol{94}3+a2*p\symbol{94}%
2+a1*p+a0))

+h0*p*(4*a4*q\symbol{94}3+3*b3*q\symbol{94}2+2*b2*q+b1)/(sqrt(a4*q\symbol{94}%
4+b3*q\symbol{94}3+b2*q\symbol{94}2+b1*q+b0)))/((p\symbol{94}3*b3*A0+p%
\symbol{94}2*A3+A1+A2*p)/(sqrt(a4*p\symbol{94}4+a3*p\symbol{94}3

+a2*p\symbol{94}2+a1*p+a0))+(A0*q\symbol{94}3*a3+A3*q\symbol{94}%
2+A4+A5*q)/(sqrt(a4*q\symbol{94}4+b3*q\symbol{94}3+b2*q\symbol{94}%
2+b1*q+b0))+A0*(4*a4*p\symbol{94}3+3*a3*p\symbol{94}2+2*a2*p+a1)

*q/(sqrt(a4*p\symbol{94}4+a3*p\symbol{94}3+a2*p\symbol{94}%
2+a1*p+a0))+A0*p*(4*a4*q\symbol{94}3+3*b3*q\symbol{94}2+2*b2*q+b1)/(sqrt(a4*q%
\symbol{94}4+b3*q\symbol{94}3+b2*q\symbol{94}2+b1*q+b0))):

a4:=1;b3:=0;b1:=0;b2:=-2;b0:=1;q:=cos(phi);qt:=-sin(phi)*ft;p:=x;pt:=xt;

a0 := -4*r*c0+d\symbol{94}2;

a1 := -4*c1*r;

a2 := -4*c2*r+2*d+4*alpha\symbol{94}2*r\symbol{94}2;

a3:=-4*r*alpha;

A0:=0;

A1:=d/r;

A2:=0;

A3:=1/r;

A4 := -1/r;

A5:=0;

h0 := 1/2*k/r;

h1 := -1/8*(4*l*a2-16*n*r-l*a3\symbol{94}2)/r;

h2 := -1/2*(-4*m*r+l*a3)/r;

h3 := 1/8*(-l*a3\symbol{94}2+4*l*a2+8*r*h1-16*n*r-8*l*d)/r/d;

h4 := -h3;

h5 := -1/2*a3*k/r;

L:=map(simplify,L,symbolic):

simplify(L-Lv);

\end{document}